\documentclass{article}
\usepackage{spconf,amsmath,graphicx, amssymb, hyperref, amsthm, dsfont, amsfonts}
\usepackage{algorithm,algorithmic}
\usepackage{multirow}
\usepackage{multicol}

\long\def\comment#1{}

\newfont{\bbb}{msbm10 scaled 700}

\newfont{\bb}{msbm10 scaled 1000}

\newcommand{\RR}{\mbox{\bb R}}

\newcommand{\vv}{{\bf v}}
\newcommand{\xv}{{\bf x}}
\newcommand{\yv}{{\bf y}}

\newcommand{\Am}{{\bf A}}
\newcommand{\Bm}{{\bf B}}
\newcommand{\Cm}{{\bf C}}

\newcommand{\Fm}{{\bf F}}

\newcommand{\Pm}{{\bf P}}
\newcommand{\Qm}{{\bf Q}}

\newcommand{\Um}{{\bf U}}

\newcommand{\Vm}{{\bf V}}

\newcommand{\Lam}{{\bf \Lambda}}

\newcommand{\Gc}{{\cal G}}

\title{Signal Variation Metrics and Graph Fourier Transforms for Directed Graphs}

\name{Laura Shimabukuro, Antonio Ortega}
\address{University of Southern California, Los Angeles, USA}

\begin{document}
\ninept
\maketitle
\begin{abstract}
In this paper we consider the problem of constructing graph Fourier transforms (GFTs) for directed graphs (digraphs), with a focus on developing multiple GFT designs 
that can capture different types of variation over the digraph node-domain. 
 Specifically, for any given digraph we propose three GFT designs based on the polar decomposition. Our method is closely related to existing polar decomposition based GFT designs, but with added interpretability in the digraph node-domain.
Each of our proposed digraph GFTs has a clear node domain variation interpretation, so that one or more of the GFTs can be used to extract different insights from available graph signals. 
We demonstrate the benefits of our approach experimentally using M-block cyclic graphs, showing that the diffusion of signals on the graph leads to changes in the spectrum obtained from our proposed GFTs, but cannot be observed with the conventional GFT definition. 

\end{abstract}

\section{Introduction}
The development of graph signal processing (GSP) tools for directed graphs is largely dependent on the construction of a suitable graph Fourier transform (GFT) that takes the directionality of the graph into account \cite{Ortega2020GSP,Sandryhaila_2013_2,ortega8347162}. 
While multiple GFT designs for digraphs have been proposed, these are predominantly focused on circumventing \cite{Sardellitti,Shafipour_2019,Chen_2022} or directly addressing \cite{Domingos_2020} the numerical instability and lack of orthogonality of the original digraph GFT, which is based on the eigendecomposition of the adjacency matrix or on the Jordan 
canonical if the adjacency matrix is not diagonalizable \cite{Sandryhaila_2013_2}. 
In each of these existing methods a specific GFT is proposed, such that its bases are well suited to capture a \textit{single} definition of total variation.

Given the wide variety of digraph structures, from directed acyclic graphs (DAGs) to M-block cyclic graphs (M-BCGs), as well as the different types of signals that can observed on any given graph,  
the main premise of our work is that  
 defining \textit{only one measure of total variation for a given digraph is unnecessarily restrictive}.  
 Specifically, 
 using the same variation for any digraph ignores that (i) certain signal variations may appear more naturally for some graph topologies and (ii)  
 there can be multiple ways for the edge weights and directionality to influence signal variation over a given digraph.
Indeed, graph signals can be related in different ways to the underlying network, with similar graph structures leading to very different types of graph signals. For example, the complex upstream and downstream propagation of traffic congestion on a highway is very different from the passive diffusion of pollution on a river network, even though both processes have directional flow constrained to branching networks \cite{particle_traffic}. 
Thus, even if both networks appear to be similar, different definitions of total variation
may be preferable in each case.


The main novelty of our work is to propose the use of multiple digraph GFTs, with their corresponding variations.
Thus, rather than defining a single definition of total variation for all digraphs, we propose to use simultaneously  multiple total variation measures. 
%
%
Starting from the polar decomposition \cite{MHASKAR2018611}, we propose three distinct bases that each capture different modes of variation over a given digraph.
Each of these distinct modes of variation provide different information about signal variation over the digraph structure.  
Applying them separately  increases the flexibility of digraph frequency analysis, and can lead to separable filtering, where only components corresponding to some types of variation are attenuated, or where different levels of attenuation are applied to each component.  
Our proposed multi-GFT digraph frequency analysis aims to develop  tools that are better adapted to account for the different dynamics of graph processes as compared to existing single-GFT methods. We also establish connections between our proposed GFTs and symmetrization strategies previously proposed in the literature. 

\textbf{Related Work:} 
The first digraph GFT definition proposed in \cite{Sandryhaila_2013_2} computes the Fourier basis as the matrix of eigenvectors of the adjacency matrix, with a frequency ordering determined by a total variation measure based on the adjacency matrix. Alternatively, the Jordan canonical form is used if the adjacency matrix cannot be diagonalized, which is often the case with directed graphs that have sinks and sources \cite{barrufet2021orthogonal}.  Subsequent work in \cite{Domingos_2020} addresses the numerical instability caused by the use of the Jordan canonical form. Other definitions \cite{7746675,Chen_2022} use the Jordan canonical form \cite{7746675} and the singular value decomposition (SVD) of the directed Laplacian \cite{Chen_2022} to construct the GFT basis. 
The methods proposed in \cite{Sardellitti,Shafipour_2019} compute their respective GFT bases through optimization methods under orthogonality constraints. The  Hermitian Laplacian based GFT constructions \cite{zhang2021magnet}, \cite{Furutani2019GraphSP} capture edge directionality through the use of complex edge weights, resulting in a unitary GFT basis. \cite{SEVI2023390} defines the GFT basis using eigenvectors of the digraph random walk operator. The method proposed in \cite{9325908} requires minor modifications of the digraph basis to ensure the diagonalizability of the adjacency matrix.
All of these papers propose a single GFT construction for a given graph, 
which corresponds to a single notion of how the directed edges of the digraph contribute to the general measure of directed signal variation.

Closest to our approach, Mhaskar \cite{MHASKAR2018611} proposed the use of the polar decomposition of the digraph adjacency matrix to develop a harmonic analysis framework on digraphs. Mhaskar's framework centers around the use of the eigenvectors of the Hermitian, positive semi-definite (PSD) factor from the left polar decomposition (LPD) to construct the GFT basis. Another closely related work \cite{unitaryshift} defines a GFT basis using the eigenvectors of the unitary factor from the polar decomposition. 
Though our proposed method makes use of both of these factors, we provide a node-domain interpretation of the variation information captured by each of these factors, which has not been provided in existing work, and use the node-domain interpretation to motivate the construction and simultaneous use of the GFT bases from all three factors obtained from both the right and left polar decompositions.
%
%

The rest of the paper is structured as follows. In Section 2 we define the three GFT bases and provide a node-domain interpretation of each of the bases by using the connection between the polar decomposition and well-known symmetrization schemes for 
digraphs \cite{symmetrizations}. In Section 3, we demonstrate the different variation information captured by each of the bases on M-block cyclic graphs.

    \begin{figure}[h]
	\centering
	\includegraphics[width=\linewidth]{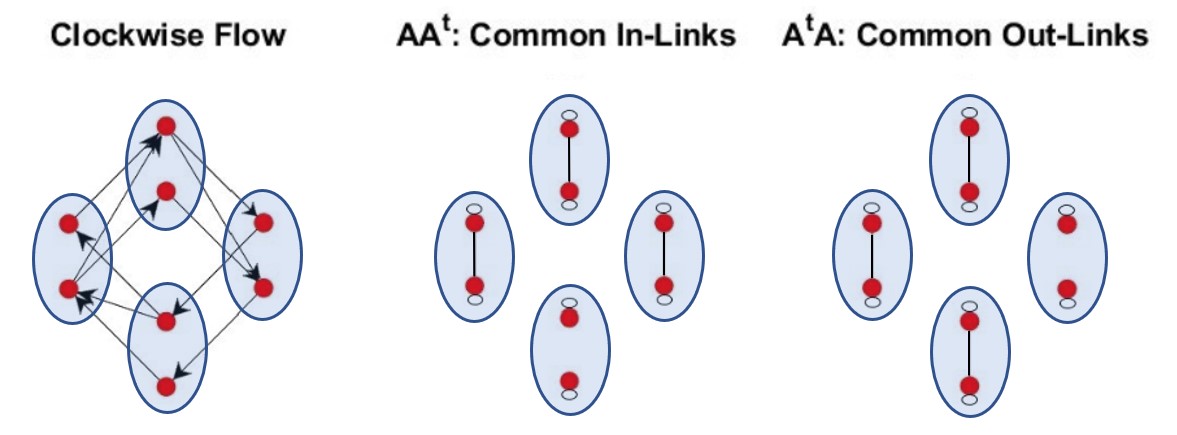}
	\caption{Example of an original digraph (left) and 
 symmetrizations $\Bm_{\text{in}} = \Am\Am^\top$  and $\Cm_{\text{out}} = \Am^\top\Am$, which can be viewed as the adjacency matrices of graphs that are composed of edges connecting nodes that share common in-links (middle) and common out-links (right) on the original graph. }
	\label{fig:symm_examples}
    \end{figure}

\section{Polar decomposition and variation}


A graph $\Gc(V,E)$ is defined by  a set of nodes $V$ and edges $E$, where an edge $e_{ij}$ represents a link from node $j$ to node $i$. A weighted graph has edges $e_{ij}$ with real positive weights $a_{ij}$.
The adjacency matrix $\Am$, an $N\times N$ square matrix, 
has $ij$ entry $a_{ij}$ corresponding to the edge weight from node $j$ to node $i$. 
Graph signals are real vectors $\xv\in \RR^N$, where the entry $x(i)$ is the real scalar corresponding to the signal corresponding to node $i$.


\subsection{Polar Decomposition}

Mhaskar \cite{MHASKAR2018611} proposed the use of the left polar decomposition (LPD), obtained from the singular value decomposition (SVD), for developing a harmonic analysis framework for digraphs, with a frequency ordering defined with respect to the singular values of the  adjacency matrix $\Am$. 
The use of the polar decomposition is key to our multiple GFT constructions because  any arbitrary $\Am$ can be factored into interpretable matrix factors.

Given an arbitrary real adjacency matrix $\Am$, its SVD is given by $\Am = \Um\Sigma\Vm^\top$, where $\Um$ and $\Vm$ are the unitary matrices of right and left singular vectors, 
respectively, and $\Sigma$ is the diagonal matrix of singular values.  
Then, the LPD is given by $\Am = \Pm\Qm$, where $\Pm$ is a positive semi-definite (PSD) Hermitian matrix and $\Qm$ is a unitary matrix, while  
the right polar decomposition (RPD) is given by $\Am = \Qm\Fm$, where $\Fm$ is a PSD Hermitian matrix and $\Qm$ is the same as in the LPD. 
LPD and RPD can be derived  from the SVD as:  
\begin{equation}
    \Am = \Um\Sigma\Vm^\top = \Um\Sigma\Um^\top\Um\Vm^\top = (\Um\Sigma\Um^\top)(\Um\Vm^\top) = \Pm\Qm, 
\end{equation}
\begin{equation}
\Am = \Um\Sigma\Vm^\top = \Um\Vm^\top\Vm\Sigma\Vm^\top = (\Um\Vm^\top)(\Vm\Sigma\Vm^\top) = \Qm\Fm.
\end{equation}

The polar decomposition, through the SVD, is directly related to the symmetrizations, $\Bm_{\text{in}} = \Am\Am^\top$ and $\Cm_{\text{out}} = \Am^\top\Am$, as can be seen from their eigendecompositions: 
\begin{equation}
\Bm_{\text{in}} = \Vm_{b}\Lam_{b}\Vm_{b}^\top = \Um\Sigma^2\Um^\top,
\end{equation} 
\begin{equation}
\Cm_{\text{out}} = \Vm_{c}\Lam_{c}\Vm_{c}^\top = \Vm\Sigma^2\Vm^\top.
\end{equation} 
The bibliographic coupling matrix, $\Bm_{\text{in}} = \Am\Am^\top$, can be viewed as the adjacency matrix of a (potentially disconnected) graph with edges connecting nodes that share common in-links on the original $\Am$ (see Fig.~\ref{fig:symm_examples}). 
The entries of $\Bm_{\text{in}}$ can be written as:
\begin{equation}
\label{eq:edges-B}
b_{ij} = \sum_ka_{ik}a_{jk}.
\end{equation}
The co-citation matrix $\Cm_{\text{out}} = \Am^\top\Am$ can be viewed as the adjacency matrix of a (potentially disconnected) graph with edges connecting nodes that share common out-links (see Fig.~\ref{fig:symm_examples}) \cite{symmetrizations}. The entries of $\Cm_{\text{out}}$ can be written as:
\begin{equation}
\label{eq:edges-C}
c_{ij} = \sum_ka_{ki}a_{kj}. 
\end{equation}
Both of these symmetrizations have been shown to capture important signal variation information in a graph signal processing context through the bibliometric symmetrization, $\Bm_{\text{in}}+\Cm_{\text{out}}$ \cite{mdp}.

In what follows, we  use the relation of the singular vectors and singular values to the symmetrizations, $\Bm_{\text{in}}$ and $\Cm_{\text{out}}$, as well as the edge weight definitions of \eqref{eq:edges-B} and \eqref{eq:edges-C} to interpret the node-domain variation captured by bases constructed from the LPD and RPD.

\subsection{Symmetrization-based variation analysis}
The clear node-domain interpretations  of the symmetrizations with respect to the original adjacency $\Am$ lead to several distinct variation modes on $\Am$. 
Quantifying the smoothness of signals on $\Bm_{\text{in}}$ and $\Cm_{\text{out}}$ using $\xv^\top\Bm_{\text{in}}\xv$ and $\xv^\top\Cm_{\text{out}}\xv$, respectively, leads to a measure of signal smoothness with respect to indirect node connections on $\Am$. Note that the variation on $\Bm_{\text{in}}$
\begin{equation}
\xv^\top\Bm_{\text{in}}\xv = \sum_ix(i)\sum_{j\in N_{\Bm}(i)}b_{ij}x(j) = 2\sum_{i\sim j}b_{ij}x(i)x(j)
\end{equation}
is maximized for the eigenvector $\vv_{b,1}$ corresponding to the largest eigenvalue $\lambda_{b,1}$. $\vv_{b,1}$ has all non-negative entries (if $\Bm_{\text{in}}$ is disconnected, there will be multiple eigenvectors corresponding to the largest eigenvalue $\lambda_{b,1}$, with strictly positive values corresponding to each disconnected component in $\Bm_{\text{in}}$). Therefore, the smoothest eigenvector $\vv_{b,1}$ for $\Bm_{\text{in}}$ is a signal on $\Am$ that has low variation over nodes that share common in-links (Fig.~\ref{fig:symm_examples}). The same reasoning shows that the ``highest frequency" eigenvector $\vv_{b,N}$, corresponding to the smallest $\lambda_{b,N}$, is a graph signal with high variation over nodes that share common in-links. The full set of eigenvectors of $\Bm_{\text{in}}$ forms an orthonormal basis that captures different levels of signal variation over nodes that share common in-links. 
Similarly, the eigenvectors of $\Cm_{\text{out}}=\Am^\top\Am$ form an  orthonormal basis that captures different levels of signal variation over nodes that share common out-links (Fig.~\ref{fig:symm_examples}).
We next use the interpretation of variations corresponding to symmetrizations to describe the properties of the GFT bases obtained from the LPD and RPD factors.

    \begin{figure}[t]
	\centering
	\includegraphics[width=\linewidth]{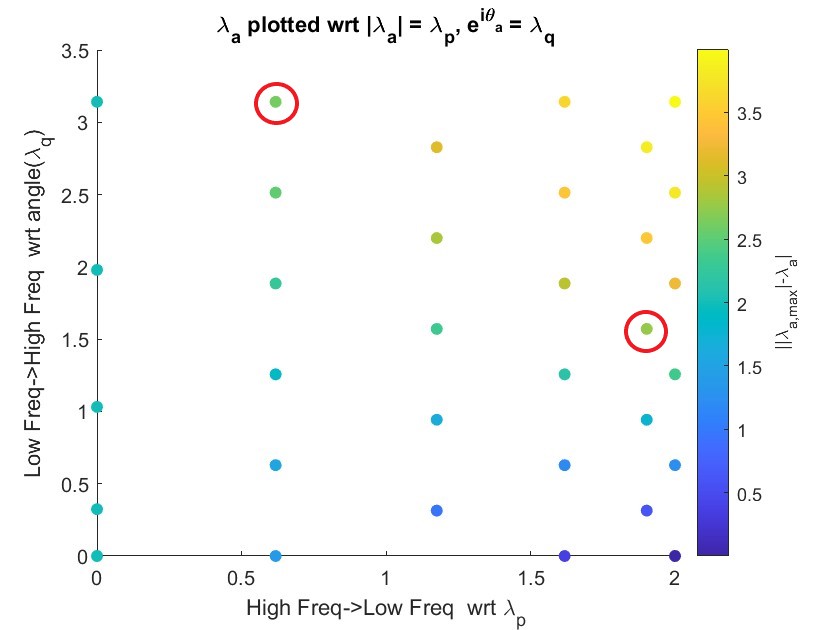}
	\caption{To illustrate the separable node-domain interpretation of the general single-mode frequency ordering based on the adjacency TV ordering \cite{Sandryhaila_2013_2}, the eigenvalues $\lambda_a$ of a N=100 directed torus graph (normal $\Am$) are plotted. As $\Am$ is normal, the direct relations $|\lambda_a| = \lambda_p$ and $e^{i\theta_a} = \lambda_q$ provide a clear node-domain interpretation of the $\lambda_a$ locations on the unit disc. Colors indicate distance of each $\lambda_a$ to $|\lambda_{a,max}|$. The two eigenvalues circled in red correspond to eigenvectors of $\Am$ with very similar total variation (TV=2.62 and TV=2.76) with respect to the total variation definition in \cite{Sandryhaila_2013_2}. However it is clear that both eigenvectors have very different variation w.r.t the spectra of $\Pm$ and $\Qm$ (i.e. very different indirect variation and inflow variation).}
	\label{fig:freq_support}
    \end{figure}

\subsection{Polar Decomposition GFT Basis Definitions}
The eigendecompositions of the three factors ($\Pm$, $\Fm$, and $\Qm$) from the LPD ($\Am=\Pm\Qm$) and RPD ($\Am=\Qm\Fm$) provide three distinct bases.
We first state the  definitions of these bases and their respective variation measures, before providing a more in-depth analysis of their node-domain variation interpretations in the subsequent section. 

\textbf{Common In-Link Basis}: Let $\Pm=\Vm_p\Lam_p\Vm_p^\top$, where the columns of $\Vm_p$ consist of the orthonormal eigenvectors of $\Pm$, and $\Lam_p$ is the diagonal matrix of non-negative eigenvalues. The set of eigenvectors $\Vm_p$ form an orthonormal GFT basis capturing variation over nodes with \textbf{common in-links} on $\Am$. The eigenvector variation ordering is determined by $\vv_p^\top\Pm\vv_p$.

\textbf{Common Out-Link Basis}: Let $\Fm=\Vm_f\Lam_f\Vm_f^\top$, where the columns of $\Vm_f$ consist of the orthonormal eigenvectors of $\Fm$, and $\Lam_f$ is the diagonal matrix of non-negative eigenvalues. The set of eigenvectors $\Vm_f$ form an orthonormal GFT basis capturing variation over nodes with \textbf{common out-links} on $\Am$. The eigenvector variation ordering is determined by $\vv_f^\top\Fm\vv_f$.

\textbf{In-Flow Basis}: Let $\Qm=\Vm_q\Lam_q\Vm_q^H$, where the columns of $\Vm_q$ consist of the unitary eigenvectors of $\Qm$, and $\Lam_q$ is the diagonal matrix of unit modulus eigenvalues. The set of eigenvectors $\Vm_q$ form a unitary GFT basis capturing variation over nodes connected by \textbf{in-flow links} on $\Am$. The eigenvector variation ordering is determined by $||\vv_q-\Qm\vv_q||_1$.

\subsection{Variation Interpretations of the Bases}

We propose to use the three GFTs ($\Vm_p, \Vm_f, \Vm_q$)  in parallel to  analyze multiple types of signal variations  on arbitrary digraphs. 

The bases $\Vm_p$ and $\Vm_f$ both capture what we refer to as \textbf{indirect variation}. Note that $\Pm=\sqrt{\Bm_{\text{in}}} = \sqrt{\Am\Am^\top}$ and $\Fm=\sqrt{\Cm_{\text{out}}}=\sqrt{\Am^\top\Am}$. As $\Bm_{\text{in}}$ and $\Cm_{\text{out}}$ are symmetric matrices and are therefore orthogonally diagonalizable, $\Bm_{\text{in}}$ and $\Cm_{\text{out}}$ share the same eigenvectors as $\Pm$ and $\Fm$, respectively. In addition, $\Pm$ and $\Fm$ share the same eigenvalues, which are the square roots of the eigenvalues of $\Bm_{\text{in}}$ and $\Cm_{\text{out}}$. Both factors can be written as $\Pm = \Vm_b\sqrt{\Lam_b}\Vm_b^\top$ and $\Fm = \Vm_c\sqrt{\Lam_c}\Vm_c^\top$ with $\Lam_b=\Lam_c$. Since $\Pm$ and $\Fm$ share the same eigenvectors as $\Bm_{\text{in}}$ and $\Cm_{\text{out}}$ with the same eigenvalue ordering (as the square root preserves their ordering), it is clear that $\Pm$ and $\Fm$ capture the same signal variation information with respect to the original $\Am$ as $\Bm_{\text{in}}$ and $\Cm_{\text{out}}$, respectively, i.e., the eigenvectors of $\Pm$ capture different levels of signal variation over nodes that share common in-links on $\Am$, and the eigenvectors of $\Fm$ capture different levels of signal variation over nodes that share common out-links on $\Am$.

The basis from $\Vm_q$ captures what we refer to as \textbf{in-flow variation}. A key property of $\Qm$ is that it is the closest unitary matrix to $\Am$ with respect to any unitarily invariant norm \cite{graph_mining}. This implies that $\Qm$ is the closest matrix to $\Am$ for which a graph multiplication is lossless ( $||\xv||_2=||\Qm\xv||_2$). Thus, the ``graph shift''  operation $\Qm\xv$ can be reversed through multiplication with $\Qm^\top$.
Similar to what we did with $\Pm$ and $\Fm$,  
we can view $\Qm$ as an alternative adjacency matrix with a direct relation to $\Am$, over which we then define our variation metric, as previously stated. 


If $\Am$ is normal, i.e., such that $\Am\Am^\top = \Am^\top\Am$, there is a direct correspondence between the magnitude and phase components of its eigenvalues $\lambda_a = |\lambda_{a}|e^{i\theta_{a}}$, and the eigenvalues of $\Pm$, $\Fm$, and $\Qm$ (for normal $\Am$, $\Pm=\Fm$). The magnitude $|\lambda_a|$ corresponds to an eigenvalue $\lambda_p$ of $\Pm$ with $\lambda_p=|\lambda_a|$. The phase $e^{i\theta_{a}}$ corresponds to an eigenvalue $\lambda_q$ of $\Qm$ with $\lambda_q=e^{i\theta_{a}}$.
Therefore, $|\lambda_{a}|$ indicates the smoothness of a signal on $\Am$ over nodes that share common-in/out links, and $e^{i\theta_{a}}$ indicates the smoothness of a signal over the in-flow connections of a graph Fig.~\ref{fig:freq_support}. This direct relation between the eigenvalues of $\Am$ and the spectrum of $\Pm$, $\Fm$, and $\Qm$ that holds for normal matrices provides a separable interpretation of the magnitude response of polynomial filters of $\Am$. 
A polynomial filter of a normal $\Am$, $p(\Am)$, assigns a specific gain with respect to each $\lambda_a$, therefore, one could implement a filtering operation in the frequency domain of $\Am$ by cascading filtering operations in the spectral domain of $\Pm$ and $\Qm$. 
Note that, as show by the circled points  Fig.~\ref{fig:freq_support}, two eigenvectors with very similar variation with respect to $\Am$, i.e., similar $||\lambda_{a,\max}|-\lambda_a|$ may have very different in-flow  ($\lambda_q$) and indirect   ($\lambda_p$) variations. This will be further shown experimentally in the case study of Section~\ref{sec:m-block}. 

However, for non-normal matrices, the lack of a direct correspondence between $\lambda_a$, $\lambda_p$, and $\lambda_q$ implies that a filter based on $\Am$ cannot be separated in terms of filters that are polynomials of $\Pm$ and $\Qm$. 
%
%
Even with this lack of correspondence for non-normal $\Am$, indirect variation and in-flow variation together allow us to interpret the eigenvalues of $\Am$ \cite{Sandryhaila_2013_2}. 
While the frequency ordering corresponding to complex $\lambda_a$ is not unique, the frequency orderings corresponding to $\lambda_p$ and $\lambda_q$ are both unique.

\section{Case study: M-Block cyclic graphs}
\label{sec:m-block}

\subsection{Interpretation} 

The difference between indirect and in-flow variation can be understood by considering an M-block cyclic graph (Fig.~\ref{fig:symm_examples}).
M-block cyclic graphs are composed of M blocks of nodes with edges connecting nodes in consecutive blocks, and zero edge connections between nodes in the same block \cite{mblock}.
Indirect variation on an M-Block cyclic graph corresponds to variation between nodes in the same block (if they share common in/out-links), whereas in-flow variation corresponds to variation between nodes in consecutive blocks. For balanced M-block cyclic graphs (same number of nodes within each block), the unitary $\Qm$ has the same M-block cyclic structure as the original $\Am$ with different edge weights. The Hermitian PSD factors, $\Pm$ and $\Fm$, have block diagonal structures (edges only within blocks) as shown in (Fig.~\ref{fig:symm_examples}).


    \begin{figure}[t]
	\centering
	\includegraphics[width=\linewidth]{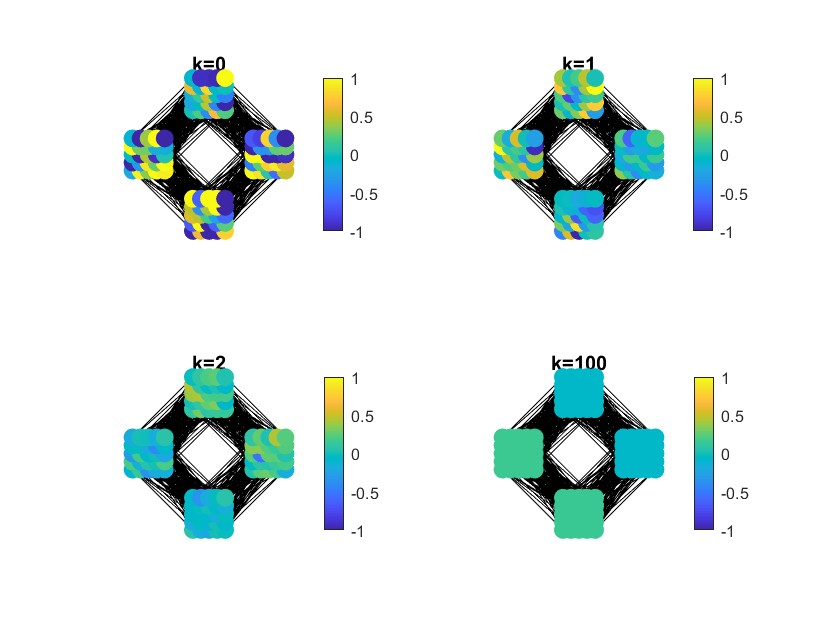}
	\caption{Signal values after k steps on an N=100 M-Block Cyclic graph, following a clockwise diffusion w.r.t the digraph edges (process flow follows the edge directions): k=0 corresponds to the iid $\mathcal{N}(0,1)$ source signals. The process eventually converges to a signal with a pattern of ``in-sync" values within subblocks (corresponding to a combination of the four eigenvectors of $\Am$ that correspond to the four largest equal magnitude eigenvalues of $\Am$ \cite{mblock}).}
	\label{fig:iid_diffusion}
\end{figure}

\begin{figure*}[t]
		\centering
		\hbox{\includegraphics[width=0.33 \linewidth]{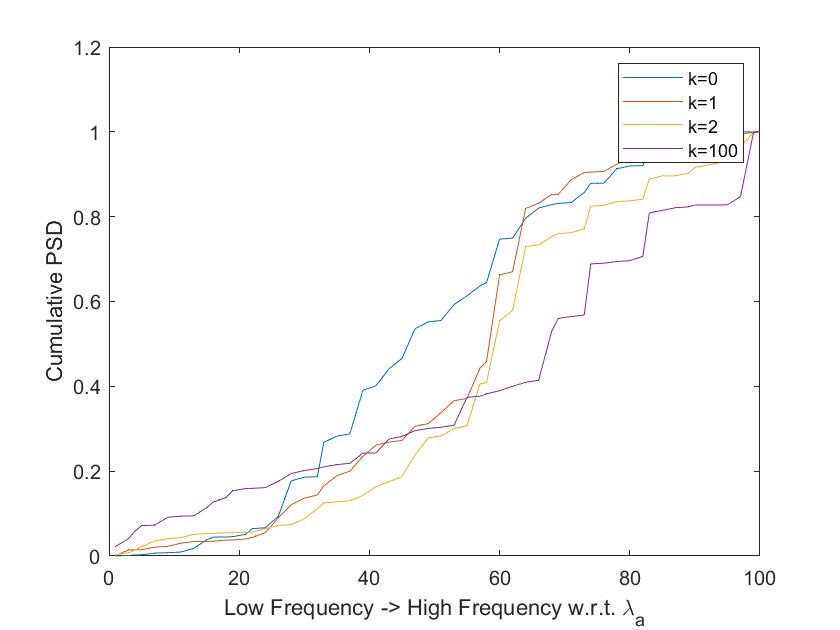}
        \includegraphics[width=0.33 \linewidth]{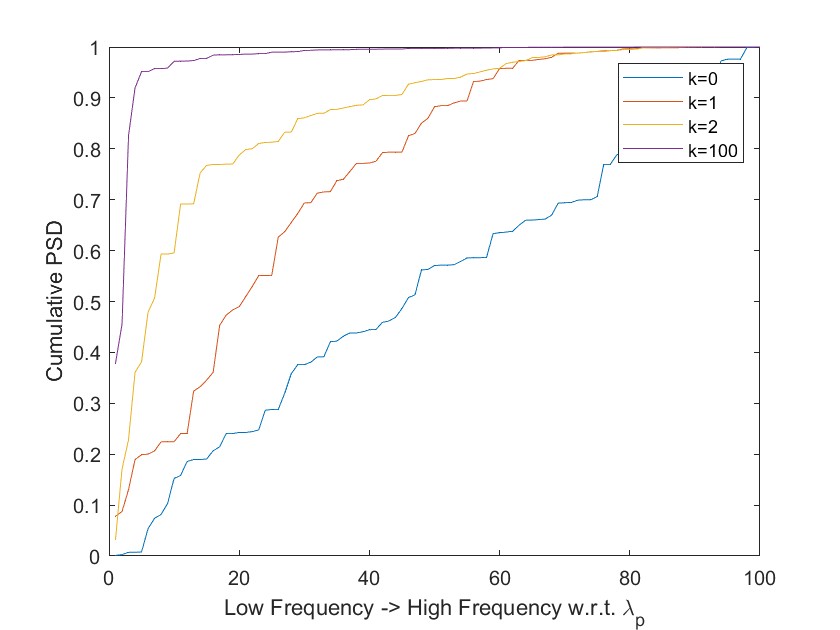}
		\includegraphics[width=0.33 \linewidth]{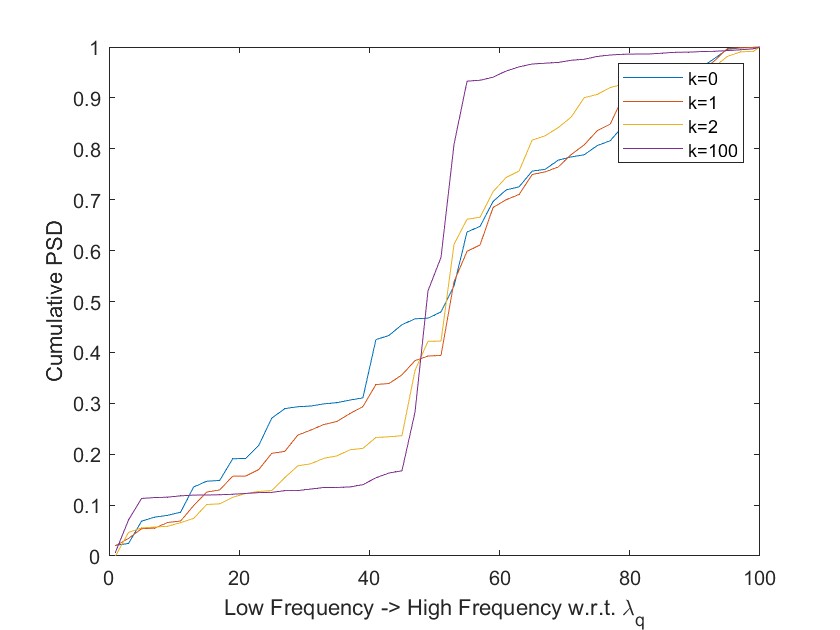}}
	\caption{At increasing steps of the diffusion process, $\yv = \Am^k\xv$, the signal begins to exhibit increasing smoothness with respect to the underlying graph structure. 
 Left: There is no clear spectrum localization for signals at any step of the process with respect to $\lambda_a$. Middle: With increasing steps, there is an increase in the spectrum localization of the signals with respect to $\lambda_p$, which corresponds to the increase in within subblock smoothness shown in Fig.~\ref{fig:iid_diffusion}. Right: With increasing steps, there is a convergence to a pattern of mid-frequency in-flow variation, as shown by the increase in signal spectrum localization with respect to $\lambda_q$. Overall, the spectrum localization exhibited by the signal with respect to the spectra of $\Pm$ and $\Qm$ highlights the observable variation patterns of the signal, while the lack of spectrum localization with respect to $\Am$'s spectrum pro.}
	\label{fig:diffusion_exp}
\end{figure*}

\subsection{M-Block cyclic graph diffusion experiments}

To demonstrate the application of separable processing of indirect and in-flow modes of variation on the same graph, we consider an M-Block cyclic graph with a non-normal adjacency $\Am$ and LPD matrices $\Pm$ and $\Qm$. 
An iid graph signal $\xv$ is generated from the standard normal distribution $\mathcal{N}(0,1)$. The signal is then diffused on the graph structure using a weighted average neighbor approach, $\yv = \Am^k\xv$, as shown in Fig.~\ref{fig:iid_diffusion}. 
The iid source signal clearly displays no smoothness properties with respect to the underlying graph structure. 
Therefore there is no spectrum localization with respect to the spectra of either $\Am$, $\Pm$, and $\Qm$ as shown in Fig.~\ref{fig:diffusion_exp}. 
With each step of the diffusion process, the signal exhibits increasing smoothness with respect to the underlying graph structure, and there is an increase in the localization of the signals' spectrum with respect to the indirect and in-flow modes of variation. 

Since indirect edge connections on M-Block Cyclic graphs are constrained to be between nodes within the same block, and the diffusion process converges to a signal that is smooth within blocks (clearly visible for the signal at the $k=100$ step), the spectral coefficients of the signal begin to localize around the low frequency components with respect to the spectrum of $\Pm$, as shown in Fig.~\ref{fig:diffusion_exp}. 
In addition to the increase in within-block smoothness at increasing steps of the process, there is also a convergence to a clean in-flow pattern of signal variation, which is most clearly visible for the signal at $k=100$. However, even the signal at $k=100$ does not result in a completely smooth signal with respect to the in-flow mode of variation, as shown by the transition to a different color at two block to block transitions. Therefore, the signal at $k=100$ shows increasing spectrum localization with respect to the spectrum of $\Qm$ around the mid-frequency range as shown in Fig.~\ref{fig:diffusion_exp}.

This example illustrates the potential advantages of separable frequency analysis using $\Pm$ and $\Qm$. Changes in the diffusion process ($k$ increases) are clearly visible in the spectral analysis of the signals as a function of the spectrum of $\Pm$ and $\Qm$  (middle and right plots in Fig.~\ref{fig:diffusion_exp}), 
there is no clear spectrum localization of the signals as $k$ grows with respect to the spectrum of $\Am$ (left Fig.~\ref{fig:diffusion_exp}).

\section{Conclusions}
We have proposed a polar decomposition based method for defining multiple graph signal variation metrics for a given digraph, with corresponding GFT bases definitions.
The guaranteed existence of the polar decomposition for all graphs ensures that these variation metrics and GFT bases always exists with unambiguous frequency orderings with respect to the corresponding spectra of $\Pm$, $\Fm$, and $\Qm$. As future work, we will investigate the use of these new variation metrics and their respective bases for designing application dependent tools that can be tuned to account for different dynamics of graph processes. 

\bibliographystyle{IEEEbib}
\bibliography{refs}

\begin{thebibliography}{10}

\bibitem{Ortega2020GSP}
Antonio Ortega,
\newblock {\em An Introduction to Graph Signal Processing},
\newblock Cambridge University Press, 2022.

\bibitem{Sandryhaila_2013_2}
Aliaksei Sandryhaila and Jose M.~F. Moura,
\newblock ``Discrete signal processing on graphs: Frequency analysis,'' 2013.

\bibitem{ortega8347162}
A.~{Ortega}, P.~{Frossard}, J.~{Kovačević}, J.~M.~F. {Moura}, and
  P.~{Vandergheynst},
\newblock ``Graph signal processing: Overview, challenges, and applications,''
\newblock {\em Proceedings of the IEEE}, vol. 106, no. 5, pp. 808--828, 2018.

\bibitem{Sardellitti}
S.~Sardellitti and S.~Barbarossa,
\newblock ``On the graph {F}ourier transform for directed graphs,''
\newblock {\em IEEE Journal of Selected Topics in Signal Processing}, vol. PP,
  Jan 2016.

\bibitem{Shafipour_2019}
Rasoul Shafipour, Ali Khodabakhsh, Gonzalo Mateos, and Evdokia Nikolova,
\newblock ``A directed graph {F}ourier transform with spread frequency
  components,''
\newblock {\em IEEE Transactions on Signal Processing}, vol. 67, no. 4, pp.
  946–960, Feb 2019.

\bibitem{Chen_2022}
Yang Chen, Cheng Cheng, and Qiyu Sun,
\newblock ``Graph {F}ourier transform based on singular value decomposition of
  directed laplacian,'' 2022.

\bibitem{Domingos_2020}
Joao Domingos and Jose M.~F. Moura,
\newblock ``Graph {F}ourier transform: A stable approximation,''
\newblock {\em {IEEE} Transactions on Signal Processing}, vol. 68, pp.
  4422--4437, 2020.

\bibitem{particle_traffic}
Dirk Helbing,
\newblock ``Traffic and related self-driven many-particle systems,''
\newblock {\em Reviews of Modern Physics}, vol. 73, 12 2000.

\bibitem{MHASKAR2018611}
H.N. Mhaskar,
\newblock ``A unified framework for harmonic analysis of functions on directed
  graphs and changing data,''
\newblock {\em Applied and Computational Harmonic Analysis}, vol. 44, no. 3,
  pp. 611--644, 2018.

\bibitem{barrufet2021orthogonal}
Julia Barrufet and Antonio Ortega,
\newblock ``Orthogonal transforms for signals on directed graphs,''
\newblock {\em arXiv preprint arXiv:2110.08364}, 2021.

\bibitem{7746675}
Rahul Singh, Abhishek Chakraborty, and B.~S. Manoj,
\newblock ``Graph {F}ourier transform based on directed laplacian,''
\newblock in {\em 2016 International Conference on Signal Processing and
  Communications (SPCOM)}, 2016, pp. 1--5.

\bibitem{zhang2021magnet}
X.~Zhang, Y.~He, N.~Brugnone, M.~Perlmutter, and M.~Hirn,
\newblock ``Magnet: A neural network for directed graphs,''
\newblock in {\em Advances in Neural Information Processing Systems},
  A.~Beygelzimer, Y.~Dauphin, P.~Liang, and J.~Wortman Vaughan, Eds., 2021.

\bibitem{Furutani2019GraphSP}
S.~Furutani, T.~Shibahara, M.~Akiyama, K.~Hato, and M.~Aida,
\newblock ``Graph signal processing for directed graphs based on the
  {H}ermitian laplacian,''
\newblock in {\em ECML/PKDD}, 2019.

\bibitem{SEVI2023390}
H.~Sevi, G.~Rilling, and P.~Borgnat,
\newblock ``Harmonic analysis on directed graphs and applications: From
  {F}ourier analysis to wavelets,''
\newblock {\em Applied and Computational Harmonic Analysis}, vol. 62, pp.
  390--440, 2023.

\bibitem{9325908}
B.~Seifert and M.~Püschel,
\newblock ``Digraph signal processing with generalized boundary conditions,''
\newblock {\em IEEE TSP}, vol. 69, pp. 1422--1437, 2021.

\bibitem{unitaryshift}
B.~S. Dees, L.~Stankovic, M.~Dakovic, A.~G. Constantinides, and D.~P. Mandic,
\newblock ``Unitary shift operators on a graph,'' 2019.

\bibitem{symmetrizations}
Venu Satuluri and Srinivasan Parthasarathy,
\newblock ``Symmetrizations for clustering directed graphs,''
\newblock in {\em Proceedings of the 14th International Conference on Extending
  Database Technology}, New York, NY, USA, 2011, EDBT/ICDT '11, p. 343–354,
  Association for Computing Machinery.

\bibitem{mdp}
Libin Liu, Arpan Chattopadhyay, and Urbashi Mitra,
\newblock ``On solving mdps with large state space: Exploitation of policy
  structures and spectral properties,''
\newblock {\em IEEE Transactions on Communications}, vol. 67, no. 6, pp.
  4151--4165, 2019.

\bibitem{graph_mining}
Georgios Drakopoulos, Eleanna Kafeza, Phivos Mylonas, and Spyros Sioutas,
\newblock ``Approximate high dimensional graph mining with matrix polar
  factorization: A twitter application,''
\newblock in {\em 2021 IEEE International Conference on Big Data (Big Data)},
  2021, pp. 4441--4449.

\bibitem{mblock}
Oguzhan Teke and P.~P. Vaidyanathan,
\newblock ``Extending classical multirate signal processing theory to
  graphs—part ii: M-channel filter banks,''
\newblock {\em IEEE Transactions on Signal Processing}, vol. 65, no. 2, pp.
  423--437, 2017.

\end{thebibliography}

\end{document}